# Characterization and Classification of Human Body Channel as a function of Excitation and Termination Modalities

Shovan Maity, Debayan Das, Baibhab Chatterjee, *Student Member, IEEE*, Shreyas Sen, *Senior Member, IEEE*
School of Electrical and Computer Engineering, Purdue University

*Abstract*— **Human Body Communication (HBC) has recently emerged as an alternative to radio frequency transmission for connecting devices on and in the human body with order(s) of magnitude lower energy. The communication between these devices can give rise to different scenarios, which can be classified as wearable-wearable, wearable-machine, machine-machine interactions. In this paper, for the first time, the human body channel characteristics is measured for a wide range of such possible scenarios (14 vs. a few in previous literature) and classified according to the form-factor of the transmitter and receiver. The effect of excitation/termination configurations on the channel loss is also explored, which helps explain the previously unexplained wide variation in HBC Channel measurements. Measurement results show that wearable-wearable interaction has the maximum loss (upto -50 dB) followed by wearable-machine and machine-machine interaction (min loss of 0.5 dB), primarily due to the small ground size of the wearable devices. Among the excitation configurations, differential excitation is suitable for small channel length whereas single ended is better for longer channel.**

## I. INTRODUCTION

Rapid advancement of semiconductor technology and process scaling over multiple decades has enabled the prolific growth of small form factor wearable devices and physiological sensors. These devices are connected to each other through radio frequency communication via air medium, and form a network of interconnected devices around the human body, commonly referred to as the Wireless Body Area Network (WBAN). Human Body Communication (HBC), which uses the human body as the communication medium, has recently emerged as an alternative to wireless media for communication among these devices due to its low power requirement and enhanced security properties. This alleviates two key technological challenges: 1) energy consumption, 2) security, for such energy constrained battery- operated devices and can enhance their lifetime significantly. One of the primary reason for energy efficiency of HBC is due to the low loss channel provided by the human body, due to the conductance property of human tissue, compared to communication via radio waves around the human body. Hence a complete understanding of the frequency characteristics of the human body channel under different communication scenarios will enable developing energy efficient HBC based circuits and systems. Previous studies on HBC channel measurement [1]–[3] have primarily focused on intra-body HBC, which characterizes the communication between two wearable devices worn by the same person. There has been wide variance in the measurement results reported in literature primarily due to the experimental setup (low impedance measurement, location of transmitter, receiver etc.) and the excitation and termination modalities. This paper investigates the different excitation termination modalities of HBC and characterizes the channel loss in such scenarios, namely wearable-wearable, wearable-machine, machine-machine interaction with differential and single ended excitation. The channel loss is measured as the ratio of received and transmitted voltage. Measurement results show that the channel characteristics is significantly different depending on the usage scenario. This knowledge can be used to design HBC based circuits depending on the application scenario and hence can be optimized to achieve maximum efficiency.

The rest of the paper is organized as follows: Section II provides the background of voltage mode HBC with capacitive termination, Section III discusses the different interaction scenarios in HBC and the different excitation and termination modalities. Section IV shows the channel loss measurements

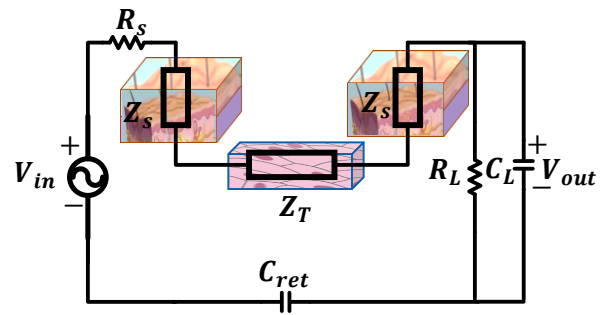

Figure 1: Simplified lumped model of human body channel for Single ended excitation and termination. The lumped capcitance at the load end $C_L$ and return path capacitance $C_{ret}$ forms a capacitive division which results in a flatband channel response enabling broadband (BB) communication.

and Section V concludes the paper summarizing the key channel loss characteristics.

## II. BACKGROUND: VOLTAGE MODE BROADBAND HBC WITH CAPACITIVE TERMINATION

The human body can be used as a broadband (BB) channel for data transmission [4], [5]. A broadband channel with 1MHz bandwidth can enable data transmission at megabits/second speed which is sufficient for applications like image, data transfer. Figure 1 shows the key components in a simplified lumped model of a HBC scenario [6]. $Z_S$ is a combination of the skin-electrode contact resistance and the skin tissue resistance and is in the order of Kilo-ohms. Once the signal is coupled to the body it is transferred through the low resistance tissue layers inside the body, whose impedance is modeled as $Z_T$ in Figure 1. The source impedance ($R_S$) is of the order of few ohms. The load and has an equivalent resistance $R_L$ and capacitance $C_L$. The load resistance is of the order of MΩs and the capacitance is in the order of tens of picofarads. In this paper we focus on the voltage transfer characteristics of the human body channel. To that end, Voltage Mode (VM) Signaling is used for communication [7], [8]. In VM signaling the measured metric at different point of the system is voltage and voltage transmission is achieved by a low output impedance source and a high input impedance load. In our current experimental set-up, a signal generator is used as a low impedance source and an oscilloscope is used as a high impedance load for measurements. High impedance termination is used since body has impedance in the range of KΩs and hence a high impedance termination in the range of MΩs provided by an oscilloscope will provide maximum voltage swing at the receiver end resulting in minimum channel loss. Capacitive termination is used at the receiver end through the capacitive load of the oscilloscope. In case of SE transmission the return path is formed by a capacitance between the transmitter and the receiver. This return path capacitance and the capacitance at the receiver end forms a capacitive voltage division, as can be seen in Figure 1. Since the HBC channel loss is determined by the ratio of return path and load capacitance and a capacitive voltage division ratio is independent of frequency, this results in a channel response with almost constant loss across all frequencies, resulting in a broadband channel. A broadband channel will enable transmission of signal through the human body directly as 1/0 bits without the need of any modulation, demodulation technique. Since broadband HBC inherently uses the complete bandwidth, it will enable the design of circuit and systems which are more energy efficient than narrowband wireless or HBC bases systems.

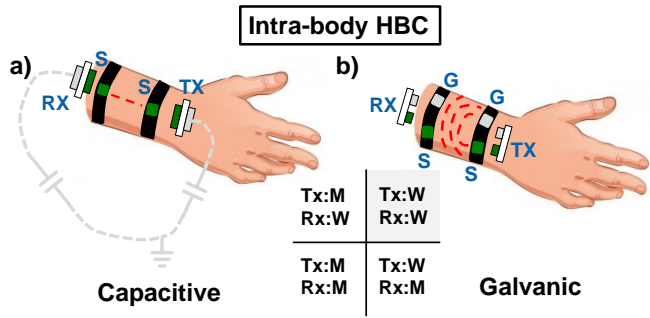

Figure 2: Intra-body HBC between two wearable devices (Tx:Wearable (W), Rx:Wearable (W)). a) Capacitive HBC with single ended(SE) excitation and termination b) Galvanic HBC with differential (DE) excitation and termination

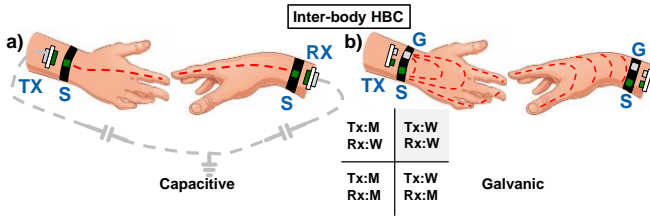

Figure 3: Inter-body HBC enabling communication through dynamic channel formed during handshake. a) Capacitive b) Differential inter-body HBC

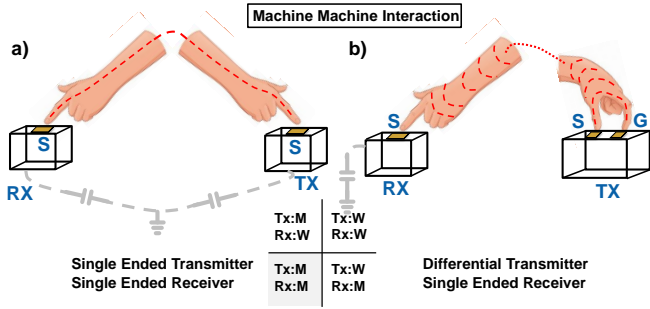

Figure 4: Communication between two off body devices through HBC. a) SE transmitter/ receiver configuration b) DE transmitter with SE receiver

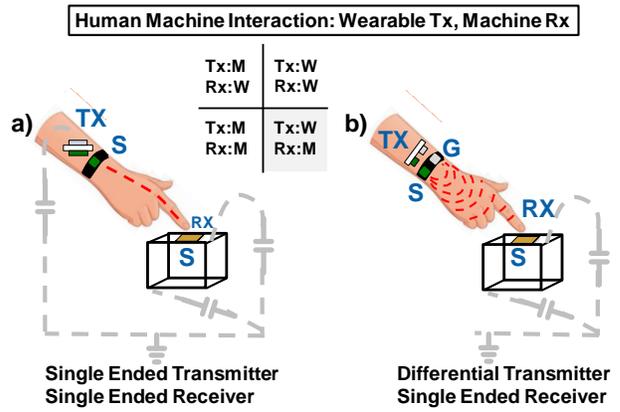

Figure 5: Interaction between a wearable transmitter and an off-body device. a) Single ended excitation provided by the transmitter, b) Differential input provided by the transmitter.

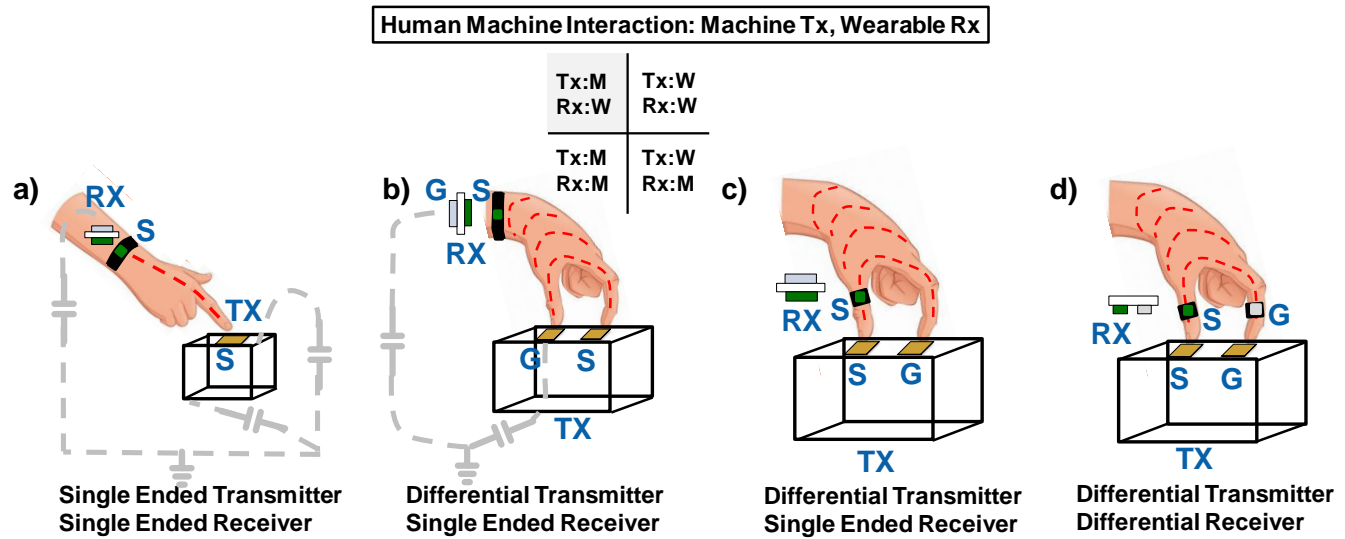

Figure 6: Different interaction scenarios between a body worn wearable as a receiver and an off-body device as a transmitter. a), b) The off-body device acts as the transmitter and the receiver is a wearable device worn on the body. c), d) Differential excitation provided by an off-body device with the receiver worn on the finger [9].

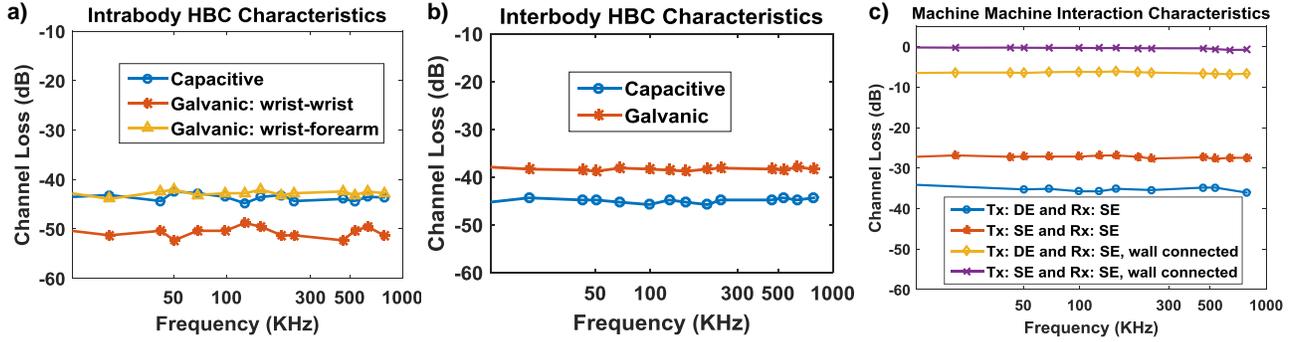

Figure 7: a) Intra-body HBC channel characteristics showing distance dependent loss for galvanic HBC, b) Inter-body HBC characteristics: galvanic HBC has lesser loss in wrist to wrist communication, c) Machine-Machine interaction characteristics: negligible loss for single ended excitation of wall connected devices.

## III. HBC CLASSIFICATION

### A. HBC Interaction Scenarios

In HBC, the transmitter couples the signal through an electrode/ a pair of electrodes into the body and it is picked up at the receiver end through electrodes after the signal gets transmitted through the human body. The human body channel can be used as a communication medium for interaction between wearable devices, sensors, as well as between a wearable device and off body electronic devices like printers, personal computers etc. This creates different interaction scenarios as discussed below:

*(1) Intra-body HBC*: Intra-body HBC refers to the communication between devices where the transmitter and receiver reside on the same person (Figure 2). Communication between a physiological sensor and a hub device in a wearable health monitoring scenario or communication between two wearable devices are examples of intra-body HBC. Previous channel measurement studies have primarily focused on characterizing the intra-body HBC channel.

*(2) Inter-Body HBC:* Inter-body HBC refers to the communication between two devices on separate human beings, where the communication channel is formed dynamically during an interaction between the two persons (Figure 3). Data transfer between smartwatches during a handshake between two persons is an example of inter-body HBC [8].

*(3) Human Machine Interaction*: HBC can enable interaction scenarios where a body worn device can communicate with a battery operated or wall connected (connected to the supply mains) machine (Figure 5, Figure 6). Transfer of images between a computer and a ring worn on the fingers of a person [9], human position tracking, secure authentication through a body worn personal key are two such possible application scenarios.

*(4) Machine-Machine Interaction:* The human body channel can also be used as a connecting medium between off body devices (Figure 4), which maybe wall connected or battery powered. Transfer of an image between a camera and a printer through touch is an example of such an interaction.

The signal transmission path is dependent on the interaction scenario and results in different amount of channel loss as will be seen in the next section.

### B. Excitation/Termination configurations

One of the other key factor in determining the channel loss is the signal excitation (at the transmitter end) and channel termination (at the receiver end) configurations. There are two primary excitation/ termination configurations:

*(1) Single Ended (SE):* In single ended excitation (Figure 2a, Figure 3a, Figure 4a, Figure 5a, Figure 6a) only the signal electrode of the transmitter is connected to the human body. The ground electrode is kept floating and the capacitive coupling between earth's ground and the ground electrode creates the return path, which enables signal transmission. Due to the capacitive return path, this is often referred to as capacitive HBC in literature.

*(2) Differential (DE):* Differential excitation is provided by connecting both the signal and ground electrode of the transmitter to the human body. This forms a closed loop within the body creating an electric field (Figure 2b, Figure 3b, Figure 4b, Figure 5b, Figure 6b,c,d), which is picked up at the receiver end. If the reception is also differential then the received voltage is the difference in potential between the two receiver electrodes. Differential excitation and termination is commonly referred to as Galvanic HBC [10] in literature.

## IV. MEASUREMENT RESULTS

Measurements are carried out to find the channel loss characteristic for the aforementioned interaction scenarios. The wearable devices are built using a Texas Instruments *TM4C123G* LaunchPad evaluation kit consisting of an ARM Cortex M4 based *TM4C123GH6PM* microcontroller. The off-body device measurements are done by putting the same device on a table to emulate the larger chassis ground. The transmitter is capable of generating signals in the frequency range of 13.3-784 KHz and the received signal is measured through an oscilloscope.

### A. Intra-body HBC

Intra-body channel characteristics is measured by applying the transmitted signal at the left wrist and measuring the received signal at two locations: left forearm and right wrist. It can be seen from Figure 7a, that the galvanic HBC loss is dependent on distance and is almost equal to capacitive HBC loss for short distances (wrist-forearm). As the distance between transmitter and receiver increases the electric field reduces at the receiver end and hence the potential difference picked up by the differential electrodes reduces, hence the channel loss increases with distance. In case of capacitive HBC, the channel loss is primarily determined by the return path capacitance, which makes it independent of transmitter receiver distance. The capacitive HBC channel characteristics is flat band as the capacitive termination at the receiver creates a capacitive division between the return path capacitance and receiver termination capacitance.

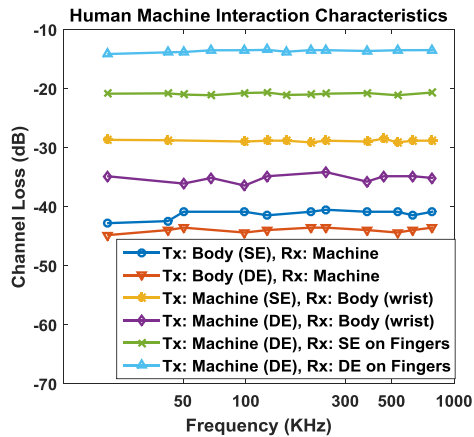

Figure 8: Channel loss characteristics for different human machine interaction scenarios, showing maximum loss with wearable transmitter

*B. Inter-body HBC*

In inter-body channel measurements, the transmitted signal is applied to the wrist of a person and the received signal is measured at the wrist of a second person during a handshake. The galvanic inter-body HBC loss is lesser than the capacitive loss (Figure 7b) due to the relatively smaller channel length between the wrists of the two persons.

*C. Human Machine Interaction*

The channel characteristics of human-machine interaction are strongly dependent on whether the wearable device is acting as the transmitter or receiver. Measurements with the wearable device acting as the transmitter show maximum loss as shown in Figure 8. Figure 6a, b shows the scenario, where the off-body machine acts as a transmitter and the received signal is measured at the wrist. Single ended excitation shows lower loss compared to differential excitation in this scenario because of the high return path capacitance due to the larger ground size of the machine. The larger ground size is also the reason for lower loss during human-machine interaction compared to a wearable-wearable interaction during intra-body or inter-body HBC. Figure 6c, d shows the scenario where the two fingers are placed at the transmitter such that it forms a closed loop and the received signal is picked up between two points within the loop. A differential reception shows minimum channel loss in this scenario. The received voltage in differential reception will increase if the distance between the receivers' electrodes increase, because there is a constant electric field between the two electrodes due to the closed path and the potential difference is proportional to the distance between the receiving electrodes.

*D. Machine-Machine Interaction*

There are two scenarios considered for channel measurement during a machine-machine interaction: 1) when the machines are battery powered, 2) when they are connected to the electric mains (wall connected). The loss depends on the excitation configuration at the transmitter end with differential excitation showing more loss compared to single ended for both wall connected or battery powered devices.(Figure 7c) Also wall connected devices show almost negligible loss as their grounds are connected through the power supply mains.

*E. Comparison between different Interactions*

Figure 9 shows a summary of all the measurements. It can be seen that when both the transmitter and receiver are wearable devices the channel loss is maximum due to small ground plane size. The channel loss of human machine interaction, which corresponds to a wearable transmitter, machine receiver (Tx: W, Rx: M) scenario, is lesser compared to wearable interactions but higher than machine-machine interaction. Finally machine-machine interaction shows the minimum loss when the devices are powered by the supply mains. Also differential excitation results in higher loss compared to single ended transmission unless the channel length is small or the received signal is picked up from within the closed loop of the transmitter.

V. CONCLUSION

This paper characterizes the human body channel under different possible interaction scenarios in a Body Area Network. Results show that the channel loss is strongly dependent on the excitation modality (differential vs single ended) and the ground sizes of the transmitter and receiver, helping explain wide discrepancies in previous measurements. Larger ground size reduces the channel loss, with supply mains connected (i.e. earth ground) machines showing minimum loss. Differential excitation shows more loss unless channel length is short or the signal is received from within the closed loop of the transmitter.

VI. ACKNOWLEDGEMENTS

This work was supported in part by the Air Force Office of Scientific Research YIP Award under Grant FA9550-17-1-0450 and in part by the National Science Foundation CRII Award under Grant CNS 1657455.

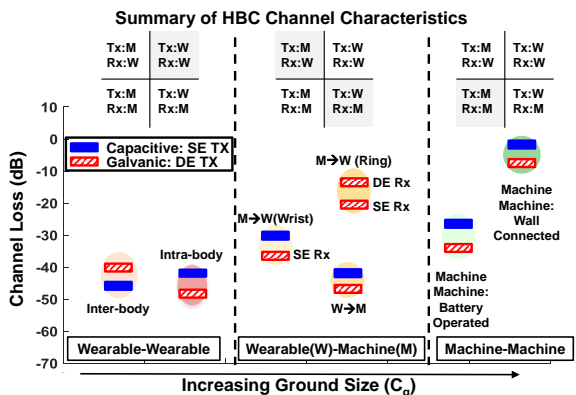

Figure 9: HBC channel measurement summary for different interaction scenarios. Larger ground size results in less loss. SE excitation (capacitive) also has lesser loss compared to DE excitation (Galvanic HBC) unless the channel length is small (wrist-forearm in experiment)

The experimental protocols involving human subjects have been approved by the Purdue University Institutional Review Board (IRB Protocol #1610018370)